\documentclass[apj]{emulateapj}
\usepackage{graphicx}
\usepackage{natbib}

\shorttitle{\emph{HST} STIS observations of the Cat's Eye Nebula}
\shortauthors{Fang et al.}

\begin{document}

\title{
\emph{HST} STIS observations of the mixing layer in the Cat's 
Eye Nebula\footnotemark[$\ast$]}

\footnotetext[$\ast$]{
Based on observations made with the NASA/ESA \emph{Hubble Space Telescope},
obtained at the Space Telescope Science Institute, which is operated by the
Association of Universities for Research in Astronomy, Inc., under NASA
contract NAS\,5-26555.
The observations are associated with program \#12509.
}

\author{
Xuan Fang$^{1}$\footnotemark[$\dagger$]\footnotemark[$\ddagger$],
Mart\'{i}n A.\ Guerrero$^{1}$,
Jes\'{u}s A.\ Toal\'{a}$^{1,2}$,
You-Hua Chu$^{2}$,
and
Robert A.\ Gruendl$^{3}$
}
\affil{
$^{1}$Instituto de Astrof\'\i sica de Andaluc\'\i a (IAA-CSIC), 
Glorieta de la Astronom\'\i a s/n, E-18008 Granada, Spain \\
$^{2}$Institute of Astronomy and Astrophysics, Academia Sinica (ASIAA), 
Taipei 10617, Taiwan \\
$^{3}$Department of Astronomy, University of Illinois, 1002 West Green 
Street, Urbana, IL 61801, USA
}
\footnotetext[$\dagger$]{Now at: 
Laboratory for Space Research, Faculty of Science, University of Hong 
Kong, Pokfulam Road, Hong Kong, China}
\footnotetext[$\ddagger$]{Also at: 
Department of Physics, University of Hong Kong, Pokfulam Road, Hong 
Kong, China}
\email{fangx@iaa.es}

\begin{abstract}

Planetary nebulae (PNe) are expected to have a $\sim$10$^5$~K 
interface layer between the $\geq$10$^6$~K inner hot bubble and the 
$\sim$10$^4$~K optical nebular shell.
The PN structure and evolution, and the X-ray emission depend critically
on the efficiency of mixing of material at this interface layer. However, 
neither its location nor its spatial extent has ever 
been determined so far.  Using high-spatial resolution \emph{HST} STIS 
spectroscopic observations of the N~{\sc v} $\lambda\lambda$1239,1243 
lines in the Cat's Eye Nebula (NGC\,6543), we have detected this 
interface layer and determined its location, extent, and physical 
properties for the first time in a PN.  We confirm that this interface 
layer, as revealed by the spatial distribution of the N~{\sc v} 
$\lambda$1239 line emission, is located between the hot bubble and the 
optical nebular shell.  We estimate a thickness of 1.5$\times10^{16}$~cm 
and an electron density of $\sim$200~cm$^{-3}$ for the mixing layer. 
With a thermal pressure of $\sim$2$\times$10$^{-8}$ dyn~cm$^{-2}$, the 
mixing layer is in pressure equilibrium with the hot bubble and ionized 
nebular rim of NGC\,6543. 

\end{abstract}

\keywords{stars: winds, outflows --- X-rays: ISM --- planetary nebulae: 
general --- planetary nebulae: individual (NGC\,6543)}

\section{Introduction} \label{sec:1}

Planetary nebulae (PNe) are formed in the final evolutionary stages of 
stars with initial masses $\leq$8--10\,$M_{\sun}$.  
As these stars evolve along the asymptotic giant branch (AGB), they 
experience successive episodes of heavy mass loss through a slow 
($v_{\infty}$ $\sim$10 km\,s$^{-1}$) wind.  
Once the stellar envelope is stripped off, the hot stellar core is 
exposed, leading to a 1000--4000 km\,s$^{-1}$ fast stellar wind 
\citep{csp85,gue13}. 
This fast wind sweeps up the slow AGB wind, which is further photoionized 
by the central star (CSPN), to form a PN \citep{kwok83,fbr90}.

In this interacting stellar winds (ISW) model, an
adiabatically-shocked hot bubble with temperatures as high as
10$^7$--10$^8$~K forms in the inner region of the PN, but this hot gas
is too tenuous ($\sim$10$^{-3}$~cm$^{-3}$) to be detected.
Nevertheless, extended X-ray emission has now been detected inside the
inner cavities of nearly 30 PNe with plasma temperatures of
1--3$\times$10$^6$~K and electron densities of 1--10~cm$^{-3}$
\citep[e.g.,][]{kast00,kast12,chu01,gue00,gue02,gue05,free14}. The
detection of X-ray-emitting hot gas in PN interiors strongly supports 
the ISW model, but the discrepancy between the observed and predicted 
physical conditions and X-ray luminosities has led to the suggestion 
that some mechanism is reducing the temperature of the hot bubble and 
raising its density.  Thermal conduction \citep[][and references 
therein]{stef08,sok94} and/or hydrodynamical instabilities 
\citep[e.g.,][]{ta14} in the wind-wind interaction zone can inject 
material into the hot bubble, creating a {\it mixing layer} of gas 
with intermediate temperatures ($\sim$10$^5$~K) between the hot bubble 
and the optical nebular shell.

As thermal conduction governs the amount of material injected into 
the hot bubble, turning it on or off in the models causes differences 
in the spatial extent and physical properties of the mixing layer.  
By gaining insights into the mixing layers in PNe, the effects of 
thermal conduction and turbulent mixing on the interior hot gas can 
be quantitatively assessed.  This in turn helps us to refine the 
models to produce more realistic predictions, which can then be 
compared with the available sample of PNe with diffuse X-ray emission 
detected \citep{free14}.  There is, however, very little observational 
information about the mixing layers.

\begin{figure}[!t]
\begin{center}
\includegraphics[width=8.4cm,angle=0]{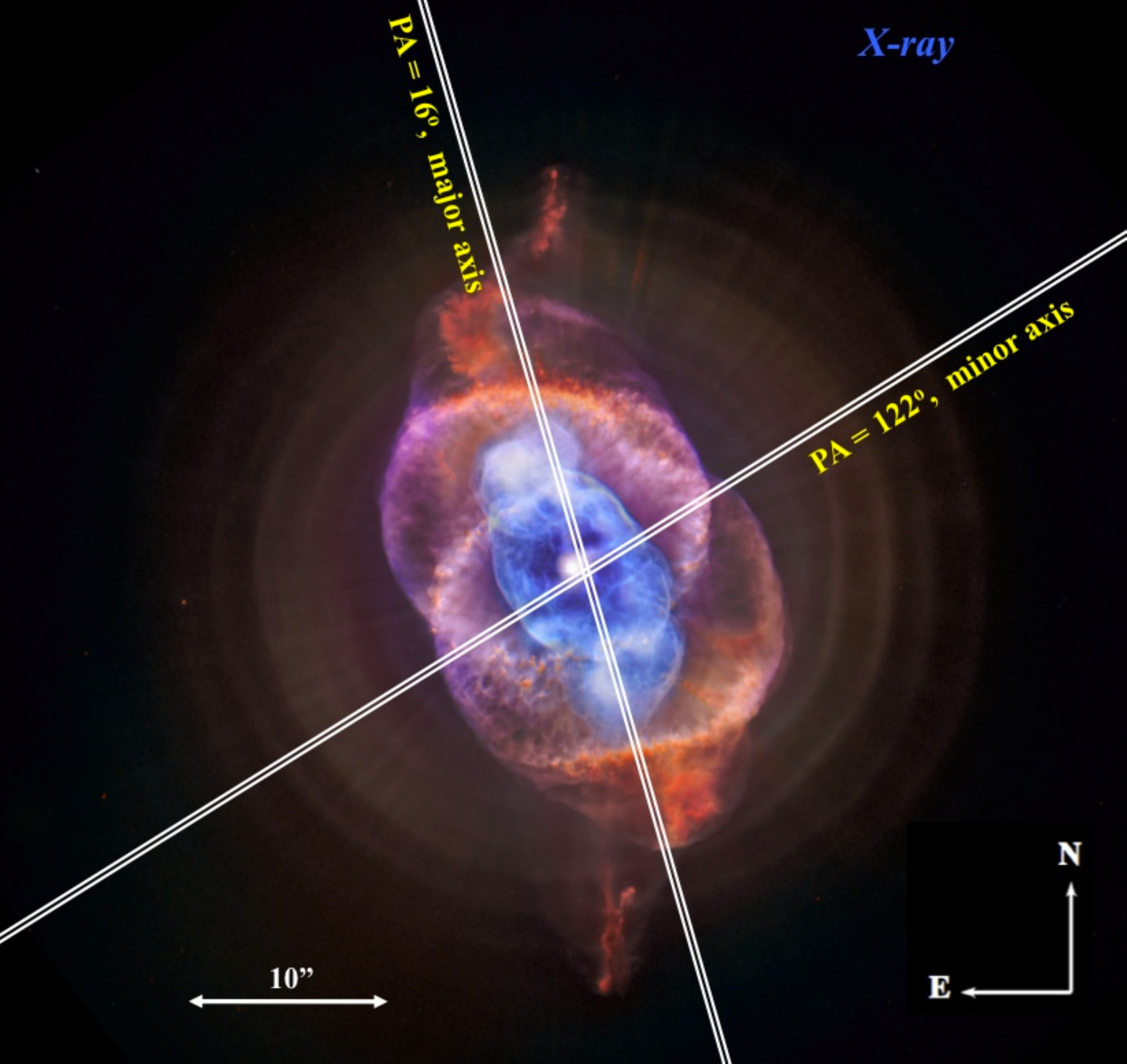}
\caption{
\emph{HST} (red, purple) and \emph{Chandra} (blue) color-composite 
image of NGC\,6543.  Image adopted from \emph{Chandra} X-ray Center 
(http://chandra.harvard.edu/photo/2008/catseye/). 
The positions of the \emph{HST} STIS 52\arcsec$\times$0\farcs2 long 
slit are marked with white lines.  At both PA =16$\degr$ and 122$\degr$, 
the slit center is offset by 0\farcs6 from the CSPN. }
\label{fig1}
\end{center}
\end{figure}

X-ray observations of NGC\,6543 (a.k.a.\ the Cat's Eye Nebula) reveal 
a physical structure qualitatively consistent with the ISW models 
\citep{chu01}.  The \emph{Chandra} image of NGC\,6543 
(Figure~\ref{fig1}) shows simple limb-brightened diffuse X-ray 
emission confined within the bright inner shell and two blisters at 
the tips of its major axis, in sharp contrast to its complex optical 
morphology \citep{bal04}, implying density enhancement near the inner 
nebular rim and evaporation of nebular material into hot interior. 
Indeed, the observed X-ray temperature 
\citep[1.7$\times$10$^{6}$~K;][]{chu01} is much lower than that 
expected for a stellar wind of $v_{\infty}$ $\sim$1400~km\,s$^{-1}$ 
\citep{pri07}.  Therefore, NGC\,6543 provides a case study of mixing 
layers in PNe.

UV lines of highly ionized species produced by thermal collisions in 
the mixing layer can be used as probes.  The most common species are 
C~{\sc iv}, N~{\sc v}, and O~{\sc vi}, whose fractional abundances 
peak at $\sim$1$\times$10$^5$, 2$\times$10$^5$ and 3$\times$10$^5$~K, 
respectively \citep{sv82}.  \emph{FUSE} detections of the O~{\sc vi} 
$\lambda\lambda$1032,1038 doublet from the mixing layers have been 
reported in several PNe \citep{ip02,ruiz13}, including NGC\,6543 
\citep{gru04}, but no spatial information could be drawn due to the 
limited angular resolution of \emph{FUSE}. 
This can only be achieved by the unique capabilities of the 
\emph{Hubble Space Telescope} (\emph{HST}).

\begin{table*}[!t]
\begin{center}
\caption{\emph{HST} STIS Observing Log}
\label{table1}
\begin{tabular}{lcllcccc}
\hline
\hline
\multicolumn{1}{c}{Slit Position} & 
\multicolumn{1}{c}{Date}          & 
\multicolumn{5}{c}{\underline{~~~~~~~~~~~~~~~~~~~~~~~~~~~~~Instrumental Configuration~~~~~~~~~~~~~~~~~~~~~~~~~~~~~}} & 
\multicolumn{1}{c}{$t_{\rm exp}$}  
\\
\multicolumn{1}{c}{} & 
\multicolumn{1}{c}{} & 
\multicolumn{1}{l}{Detector} & 
\multicolumn{1}{c}{Grating}  & 
\multicolumn{1}{c}{$\lambda_c$}    & 
\multicolumn{1}{c}{Spectral Range} & 
\multicolumn{1}{c}{Dispersion}     & 
\multicolumn{1}{c}{} \\
\multicolumn{1}{c}{} & 
\multicolumn{1}{c}{} & 
\multicolumn{1}{c}{} & 
\multicolumn{1}{c}{} & 
\multicolumn{1}{c}{({\AA})} & 
\multicolumn{1}{c}{({\AA})} & 
\multicolumn{1}{c}{({\AA}~pixel$^{-1}$)} & 
\multicolumn{1}{c}{(s)}           
\\
\hline
NGC\,6543-MINOR
               & 2012 Jul. 3  & STIS/FUV-MAMA & G140M & 1222 & 1190--1253 & 0.053 & 4570 \\
               &              & STIS/FUV-MAMA & G140M & 1550 & 1518--1581 & 0.053 & 1723 \\
               &              & STIS/CCD      & G430L & 4300 & 2652--5950 & 2.746 & 2$\times$127 \\
               &              & STIS/CCD      & G750M & 6581 & 6248--6913 & 0.554 & 2$\times$275 \\
NGC\,6543-MAJOR 
               & 2012 Oct. 21 & STIS/FUV-MAMA & G140M & 1425 & 1190--1253 & 0.053 & 4570 \\
               &              & STIS/FUV-MAMA & G140M & 1550 & 1518--1581 & 0.053 & 1723 \\
               &              & STIS/CCD      & G430L & 4300 & 2652--5950 & 2.746 & 2$\times$127 \\
               &              & STIS/CCD      & G750M & 6581 & 6248--6913 & 0.554 & 2$\times$275 \\
\hline
\end{tabular}
\end{center}
\end{table*}

In this paper, we present \emph{HST} STIS UV and optical spectroscopy 
of NGC\,6543.  In conjunction with the \emph{Chandra} X-ray images, 
these new spectra are used to successfully determine the location and 
spatial extent of mixing layer in a PN for the first time.  We 
describe the observations in Section~\ref{sec:2}, and present results 
and discussion in Section~\ref{sec:3}.  The main conclusions are 
summarized in Section~\ref{sec:4}.

\section{Observations and Data Analysis} \label{sec:2}

\emph{HST} STIS UV and optical spectroscopic observations of NGC\,6543 
(PI: M.A.\ Guerrero, GO prop.~ID 12509, Cycle~19) were carried out on 
2012 July 3 and 2012 October 21.  The observations aimed at detecting 
and tracing the spatial extent of the interface layer and comparing it 
with those of the nebular shell and hot bubble.  
The 52\arcsec$\times$0\farcs2 long slit was placed at a position angle 
(PA) of 16$\degr$ and 122$\degr$ along the major and minor axes of the 
inner nebular shell (Figure~\ref{fig1}), respectively.  The G140M 
grating and STIS/FUV-MAMA detector were used to acquire spectra of
the N~{\sc v} $\lambda\lambda$1239,1243 and C~{\sc iv} $\lambda$1548,1551 
emission lines.  The observations were performed in ACCUM mode. 
Meanwhile, the G430L and G750M gratings and STIS/CCD detector were used 
to obtain information of the [O~{\sc iii}], H$\alpha$, and [N~{\sc ii}] 
lines from the optical nebular shell.  The STIS/CCD observations were 
split into 2 exposures to allow cosmic-ray removal.  A summary of the 
STIS configurations and exposures is given in Table~\ref{table1}.  The 
spectra were reduced and calibrated with the \emph{HST} STIS pipeline.

\begin{figure}[!t]
\begin{center}
\includegraphics[width=8.7cm,angle=0]{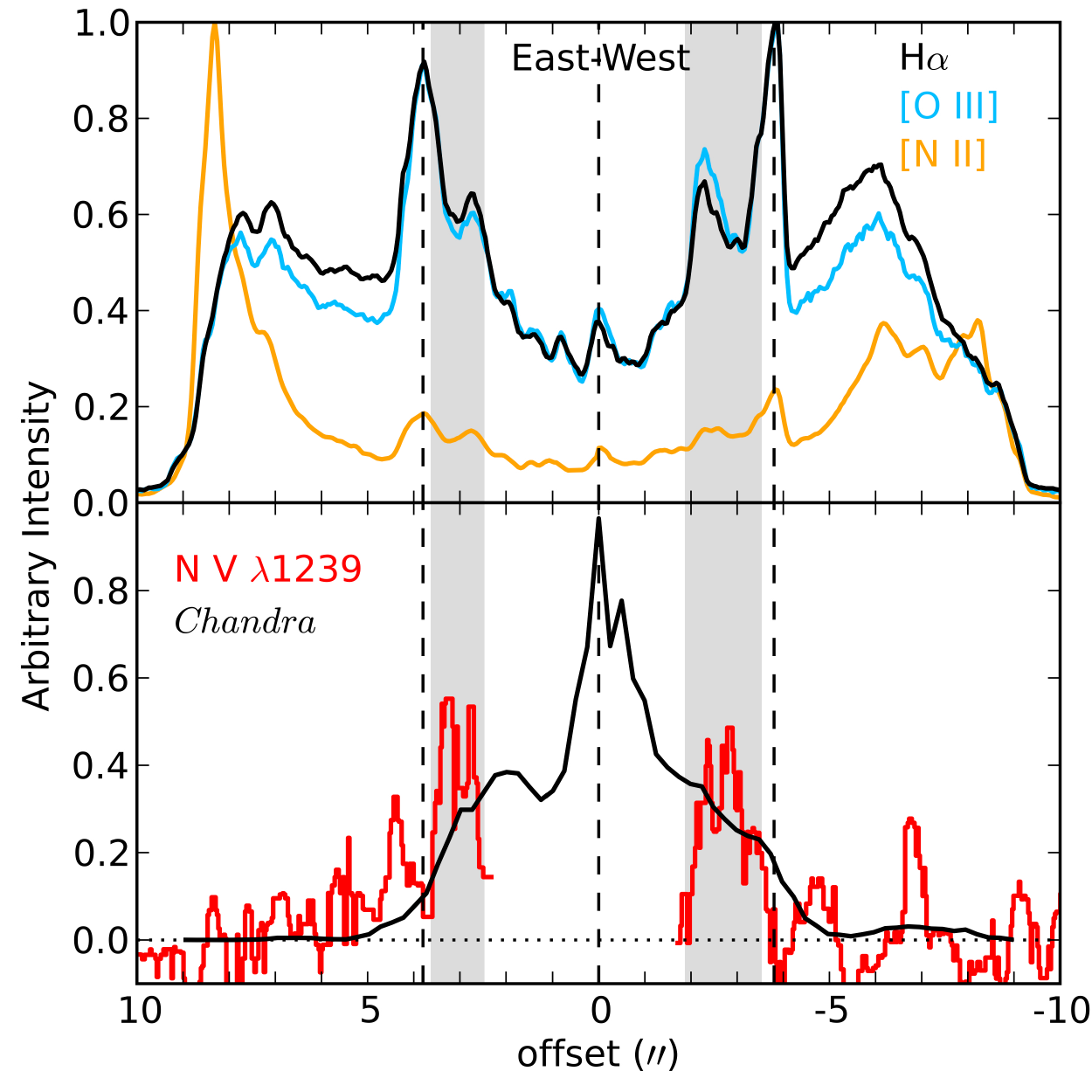}
\caption{
Spatial emission profiles of the H$\alpha$, [O~{\sc iii}] $\lambda$5007, 
and [N~{\sc ii}] $\lambda$6583 emission lines from the optical nebular 
shell (top panel), and the N~{\sc v} $\lambda$1239 UV resonance line 
from the interface layer and the \emph{Chandra} X-ray emission from 
the hot bubble (bottom panel) along the minor axis of NGC\,6543 at PA 
122\degr (Figure~\ref{fig1}).  The offset is relative to the CSPN. 
Grey shaded areas mark the positions of the interface layer, as 
indicated by the N~{\sc v} line. }
\label{fig2}
\end{center}
\end{figure}

In order to avoid the bright stellar light, the center of the long 
slit was offset 0\farcs6 from the CSPN at each slit PA.  Despite 
this offset, noticeable scattered stellar continuum spilt into the 
innermost regions of the nebular shell, reducing the detection 
sensitivity for the faint nebular emission within a region of radius 
$\simeq$1\farcs5 around the CSPN.

The 2D STIS spectra were used to extract spatial profiles of emission 
along the minor axis of the photoionized innermost nebular shell for 
the H$\alpha$, [O~{\sc iii}] $\lambda$5007, and [N~{\sc ii}] 
$\lambda$6583 lines (Figure~\ref{fig2}, top) and the 
collisionally-excited N~{\sc v} $\lambda$1239 UV line 
(Figure~\ref{fig2}, bottom).  The stellar light and nebular continuum 
were subtracted by carefully selecting spectral regions blue- and 
red-wards of the target emission lines.  Despite this effort, the 
steep stellar P-Cygni profile of the N~{\sc v} line made it impossible 
to obtain a clean spatial profile of nebular emission in the innermost 
3\arcsec\ region around the CSPN. 
This inner section of spatial profile is discarded in our analysis. 
The spatial profile of X-ray emission, as derived from the 
\emph{Chandra} observations, is added into the bottom panel of 
Figure~\ref{fig2}.  The spatial profiles along the major axis of 
the inner nebular shell of NGC\,6543 (not shown here) reveal similar 
structures, although are complicated by projection effects of 
emission from the blister features.

\begin{figure}[!t]
\begin{center}
\includegraphics[width=8.4cm,angle=0]{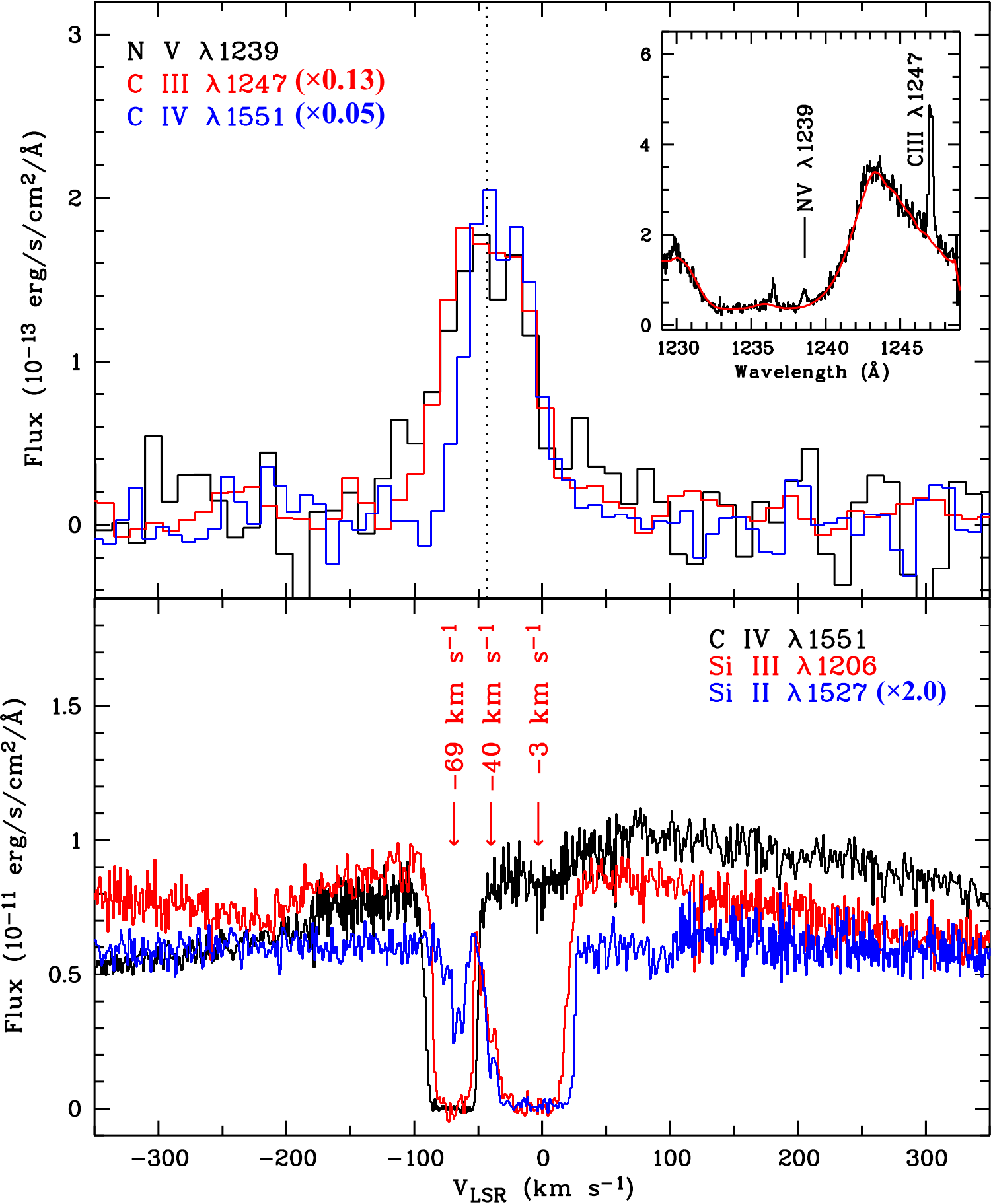}
\caption{
Top panel: 
Nebular N~{\sc v} $\lambda$1239, C~{\sc iii} $\lambda$1247, and 
C~{\sc iv} $\lambda$1551 emission lines extracted at the mixing layer 
(Figure~\ref{fig2}) along the minor axis of NGC\,6543. Wavelengths 
have been converted to the LSR velocity.  The radial velocity of 
NGC\,6543 \citep[$-$47.5~km\,s$^{-1}$;][]{ms92} is marked by the 
vertical dotted line.  Spectra are binned according to the dispersion 
(0.053\,{\AA}\,pixel$^{-1}$) of the STIS G140M grating.  Fluxes of the 
C~{\sc iii} and C~{\sc iv} lines are scaled by 0.13 and 0.05, 
respectively, to allow better comparison with the N~{\sc v} line. 
The inset shows the 1D spectrum of the mixing layer in 1229--1249\,{\AA}, 
and the red continuous curve is the P-Cygni profile of the CSPN. 
Bottom panel: 
STIS E140H spectrum of NGC\,6543's central star corrected for the LSR 
velocity showing the absorption features of C~{\sc iv} $\lambda$1551 
(black), Si~{\sc iii} $\lambda$1206 (red) and Si~{\sc ii} $\lambda$1527 
(blue).  Absorptions due to the PN shell and the interstellar gas are 
marked and velocities presented. } 
\label{fig3}
\end{center}
\end{figure}

We extracted 1D spectra from the 2D STIS UV data along the minor 
axis.  The apertures for spectral extraction were targeted at the 
location of the interface layer marked by the grey shaded regions 
in the bottom panel of Figure~\ref{fig2} that have been selected 
according to the spatial emission profile of the N~{\sc v} line. 
The spectra extracted at the east and west positions of the interface 
layer were then combined and corrected for the underlying scattered 
stellar continuum.  The profiles of the N~{\sc v} $\lambda$1239, 
C~{\sc iii} $\lambda$1247, and C~{\sc iv} $\lambda$1551 emission 
lines are presented in Figure~\ref{fig3}-top.  Wavelengths have been 
corrected for the instrumental and orbital shifts.  We then converted 
wavelengths to velocities by correcting for the local-standard-of-rest 
(LSR) velocity of the solar system ($v_{\rm LSR}$ = 16~km\,s$^{-1}$) 
towards the direction of NGC\,6543 ($l$=96\fdg5, $b$=30\fdg0).

Archival \emph{HST} STIS echelle spectra of the CSPN of NGC\,6543 
(PI: R.E.\ Williams, GO prop.~ID 9736, Cycle~12) were used to 
complement our data analysis.  The stellar spectra were obtained with 
the E140H grating, which provided a resolution of 114,000\footnote{The 
STIS Instrument Handbook, URL 
http://www.stsci.edu/hst/stis/documents/handbooks} ($\sim$3 
km\,s$^{-1}$).  Three separate settings of STIS/FUV-MAMA were used 
to cover a wavelength region 1140--1690 \AA.  The 
0\farcs2$\times$0\farcs09 slit was placed on the CSPN.  Detailed 
description of observations is given in \citet{wil08}.

\section{Results and Discussion} \label{sec:3}

\subsection{Emission Line Profiles} \label{sec:3:a}

Figure~\ref{fig3}-top shows the N~{\sc v}, C~{\sc iii} and C~{\sc iv} 
emission lines detected in our STIS G140M spectrum of NGC\,6543.  The 
N~{\sc v} $\lambda$1239 and C~{\sc iii} $\lambda$1247 lines peak at 
the radial velocity of NGC\,6543 \citep[$-$47.5 km\,s$^{-1}$;][]{ms92} 
and both have a full-width at half-maximum (FWHM) $\sim$0.33\,{\AA}. 
This line width is expected for an extended source filling the STIS 
slit width (0\farcs2) and actually matches that of the geocoronal 
Ly$\alpha$ emission line.  This spectral resolution is insufficient 
to resolve the thermal width of N~{\sc v} $\lambda$1239 produced by 
the 2$\times$10$^{5}$~K gas in the interface layer, which is estimated 
to have a FWHM $\sim$0.11\,{\AA}.  Meanwhile, both lines of the C~{\sc 
iv} $\lambda\lambda$1548,1551 doublet have a FWHM$\sim$0.31\,{\AA}, 
slightly narrower than N~{\sc v} and C~{\sc iii} (Figure~\ref{fig3}, 
top).  Moreover, its observed line center is redshifted by 0.078\,{\AA} 
($\sim$15 km\,s$^{-1}$), compared to the N~{\sc v} and C~{\sc iii} 
lines.

Inspection of our 2D STIS spectrum reveals a narrow absorption 
bluewards of the C~{\sc iv} emission.  A close look at the archival 
\emph{HST} STIS E140H spectrum of NGC\,6543's central star helps to 
clarify this issue.  The spectra of the Si~{\sc ii} $\lambda$1527, 
Si~{\sc iii} $\lambda$1206, and C~{\sc iv} $\lambda$1551 lines have 
absorption features at $-$3, $-$40 and $-$69 km\,s$^{-1}$ 
(Figure~\ref{fig3}, bottom).  The component at $-$3 km\,s$^{-1}$ is 
saturated in Si~{\sc ii} and Si~{\sc iii} but weak in C~{\sc iv}, 
whereas the much weaker absorption at $-$40 km\,s$^{-1}$ is only 
present in Si~{\sc ii} and Si~{\sc iii}.  These two absorption features 
are generally consistent in radial velocity and relative strength with 
the H~{\sc i} 21~cm emission towards the direction of NGC\,6543 
detected in the Leiden/Argentine/Bonn Galactic H~{\sc i} Survey and 
Effelsberg-Bonn H~{\sc i} Survey \citep{kal05,win16}.  Given the 
similar ionization potentials of H$^{0}$ and Si$^{+}$, these two 
absorption components can be attributed to neutral or low-excitation 
ionized interstellar gas along the direction of NGC\,6543. 
On the other hand, the absorption at $-$69 km\,s$^{-1}$ is most likely 
produced by the approaching side of the PN shell, as this velocity 
generally agrees with NGC\,6543's systemic velocity ($v_{\rm LSR}$ = 
$-$47.5~km\,s$^{-1}$) plus its expansion velocity ($\sim$16 
km\,s$^{-1}$ at the inner shell and 28 km\,s$^{-1}$ at the outer 
shell; \citealt{ms92}).  Furthermore, this component is weak and 
unsaturated in Si~{\sc ii}, saturated in Si~{\sc iii}, and heavily 
saturated in C~{\sc iv}, implying a higher excitation than that of 
the interstellar gas probed by the H~{\sc i} 21~cm surveys.  Our 
identification of these absorption features generally agrees with the 
interpretation of \emph{IUE} observations \citep{pwa84}.

The blueward absorption reduces the widths of the C~{\sc iv} emission 
lines and shifts their line centers towards the red.  This explains 
the different emission line profiles of C~{\sc iv} with respect to 
those of N~{\sc v} and C~{\sc iii} seen in Figure~\ref{fig3}-top.

\subsection{Spatial Distribution of Line Emission} \label{sec:3:b}

The spatial profiles along the minor axis of NGC\,6543 
(Figure~\ref{fig2}) reveal the location of mixing-layer gas 
originating from very different processes.  The brightest emission 
peaks in H$\alpha$ and [O~{\sc iii}] at $\simeq$3\farcs9 from the 
CSPN mark the location of the $\sim$10$^4$~K swept-up inner nebular 
shell.  The spatial profiles of the C~{\sc iii} and C~{\sc iv} lines 
are generally consistent with those of the H$\alpha$ and [O~{\sc iii}] 
lines associated with the inner shell.  On the other hand, the profile 
of the X-ray emission from the $\gtrsim$10$^6$~K hot gas shows an 
eastern peak at $\simeq$2\farcs0 and a shoulder of declining emission 
towards the west.  This irregular profile is due to the low count 
rate, but suggests that the X-ray-emitting gas is confined within 
a region with radius $\leq$3\arcsec.  These new profiles confirm 
the interpretation of \citet{chu01} that the X-ray-emitting gas is 
confined within the cool nebular shell.

Interestingly, the useful section of the spatial profile of the 
N~{\sc v} emission peaks at intermediate positions (grey shades in 
Figure~\ref{fig2}), $\simeq$3\arcsec, between the optical lines 
from the optical nebular shell and the X-ray emission from the hot 
bubble.  The N~{\sc v} ion cannot be produced by photoionization 
because the effective temperature of the CSPN of NGC\,6543 
is only $\simeq$50,000~K; thus it must be produced by thermal 
collisions at temperatures of $\sim$10$^5$~K, as expected in the 
mixing layer.

The observed spatial profile of the N~{\sc v} $\lambda$1239 emission 
line can be used to estimate the radius and thickness of mixing layer. 
Assuming a constant-emissivity cylindrical shell with radius $R$ and 
thickness ${\Delta}R$, we derived an outer radius of $\sim$3\farcs7 
and a thickness 27\% this radius (i.e., 1\farcs0).  At a distance of 
1.0$\pm$0.3~kpc \citep{reed99}, this implies a thickness of 
1.5$\times$10$^{16}$~cm.  The H$\alpha$ profile can be fit similarly 
to derive an outer radius of 4\farcs6 and a thickness of 0\farcs9, 
making its inner edge coincident with the outer edge of the mixing 
layer.

The estimated thickness of the mixing layer in NGC\,6543 can be 
compared with our numerical results \citep{ta14}.  Our post-AGB model 
with 0.633~$M_{\odot}$ predicts a mixing layer with a thickness of 
1.8$\times10^{16}$~cm by the time the hot bubble reaches a similar 
average radius as that of NGC\,6543 ($R_{\rm bubble}$ 
$\lesssim\,0.04$~pc).  This is very similar to the measured thickness. 
However, this thickness only covers 10 cells in our current models, 
and thus the mixing layer is not sufficiently sampled.  New 
high-resolution simulations are needed to make accurate predictions 
on the evolution and physical properties (spatial extent, density, 
and temperature) of the mixing layers in PNe.

\subsection{Mixing Layer Electron Density and Pressure} 
\label{sec:3:c}

We estimated the density of the mixing layer by assuming a simple 
geometry of the N~{\sc v} $\lambda$1239-emitting region: 
a cylindrical shell with an outer radius of 3\farcs7 and an inner 
radius of 2\farcs7.  
The intensity of the N~{\sc v} $\lambda$1239 line of the interface 
layer, as measured from the extracted spectrum, is 
6.3$\times$10$^{-14}$ erg~cm$^{-2}$~s$^{-1}$.  This line intensity can 
be expressed as: 
\begin{equation}
I = n_{\rm e} n_{\rm N^{4+}} h\nu 
    \frac{8.629 \times 10^{-6}}{\sqrt{T_{\rm e}}} 
    \frac{\Omega(1,2; T_{\rm e})}{g_{1}} 
    {\rm e}^{-\chi/kT_{\rm e}} 
    \frac{V}{4{\pi}d^{2}}. 
\label{eq1}
\end{equation}
Here $n_{\rm e}$ and $n_{\rm N^{4+}}$ are number densities of the 
electron and the N$^{4+}$ ion, respectively; 
$h\nu$ is the photon energy (in ergs) of the N~{\sc v} $\lambda$1239 
line; $\Omega$(1,2; $T_{\rm e}$) is the Maxwellian-averaged collision 
strength of the N$^{4+}$ 
2s\,$^{2}$S$_{1/2}$ -- 2p\,$^{2}$P$^{\rm o}_{3/2}$ transition, which 
is derived using the collision strength of N$^{4+}$ 2s\,$^{2}$S -- 
2p\,$^{2}$P$^{\rm o}$ calculated by \citet{cm83} and Equation~(3.21) 
in \citet{of06}; $g_{1}$ is the statistical weight of the lower level 
(for N$^{4+}$, $g_{1}$ = 2); $\chi$ is the excitation energy of the 
upper level (in the case of N~{\sc v} $\lambda$1239, $\chi$ = 
10.008~eV, which corresponds to 1.603$\times$10$^{-11}$ ergs); $d$ 
is the distance to NGC\,6543; $V$ is the emitting volume in cm$^{3}$.

Here $V$ can be derived from the STIS slit width (0\farcs2, 
corresponding to 3.0$\times$10$^{15}$~cm) and the cylindrical 
shell of the UV-emitting mixing layer (as assumed at the 
beginning of this section) using the equation 
\begin{equation}
V = 2\pi\Delta{x} 
      \left(
        \int_a^{R_1}\sqrt{R_1^2 - r^2}~\mathrm{d}r - 
        \int_{a}^{R_2}\sqrt{R_{2}^{2} - r^{2}}~\mathrm{d}r 
      \right)
\label{eq2}
\end{equation}
where $R_{1}$ and $R_{2}$ are the outer and inner radius of the cylindrical 
shell, respectively (5.5$\times$10$^{16}$~cm and 4.0$\times$10$^{16}$~cm 
at 1~kpc), and $\Delta{x}$ is the thickness covered by the STIS long 
slit (3.0$\times$10$^{15}$~cm); $a$ is the lower limit of the 
integration, which corresponds to 3.7$\times$10$^{16}$~cm from the 
CSPN according to the region selected for spectral extraction.  
The likely inhomogeneity of the 2$\times$10$^5$~K, N~{\sc v} 
$\lambda$1239-emitting gas due to hydrodynamical instabilities can 
be accounted for by adding a filling factor, $\epsilon$, to 
Equation~\ref{eq1}.

In the interface layer, the N$^{4+}$/H$^{+}$ ionic abundance ratio 
was assumed to be close to the nebular nitrogen abundance (N/H), 
which is 2.30$\times$10$^{-4}$ \citep{ber03}.  This nebular abundance 
is consistent with the stellar wind abundance 
\citep[2.29$\pm$0.53$\times$10$^{-4}$;][]{geor08}.  Combining 
Equations~\ref{eq1} and \ref{eq2} and by introducing a filling factor 
$\epsilon$, we deduced an expression of the electron density as a 
function of temperature, $n_{\rm e}$ = 
6.1\,$\epsilon^{-1/2}$\,$T_{\rm e}^{1/4}$\,$\exp$(5.808$\times$10$^{4}$/$T_{\rm e}$). 
This function shows that when the temperature varies in the range 
1$\times$10$^5$--3$\times$10$^5$~K, the electron density is always 
close to $\sim$180\,$\epsilon^{-1/2}$~cm$^{-3}$ for the interface 
layer.

This density and the adopted temperature of 2$\times$10$^5$~K imply 
a thermal pressure of $\sim$2$\times$10$^{-8}$\,$\epsilon^{-1/2}$ 
dyn~cm$^{-2}$ in the mixing layer, which agrees with the pressure 
of the hot bubble and the ionized swept-up shell \citep{gru04}, 
probably implying the filling factor $\epsilon\sim$1. 

\section{Conclusions} \label{sec:4}

We present high-spatial resolution \emph{HST} STIS UV and optical 
spectroscopy of the Cat's Eye Nebula (NGC\,6543).  Our STIS 
observations enabled the first view of the spatial distribution of 
the mixing-layer gas.  This mixing layer, probed by the 
collisionally-ionized N~{\sc v} UV emission line, is located exactly 
between the optical nebular rim and the X-ray-emitting hot bubble as 
previously detected by \emph{Chandra}.

We estimate a thickness of 1.5$\times10^{16}$~cm for the mixing layer, 
which is consistent with predictions of our 2D radiation-hydrodynamic 
simulations of the hot bubbles in PNe \citep{ta14}.  The estimated 
electron density and thermal pressure of this layer are found to be 
$\sim$180\,$\epsilon^{-1/2}$~cm$^{-3}$ and 
$\sim$2$\times$10$^{-8}$\,$\epsilon^{-1/2}$~dyn\,cm$^{-2}$, 
respectively, assuming a cylindrical shell of the 2$\times10^{5}$~K, 
UV-emitting gas.  This thermal pressure agrees with that in the hot 
bubble and ionized nebular rim of NGC\,6543, suggesting hydrodynamical 
equilibrium.  New higher-resolution radiation-hydrodynamic numerical 
simulations will be carried out to investigate the evolution and 
properties of the mixing layer in young PNe (Toal\'{a} \& Arthur, in 
preparation).

It is worth mentioning that the physical configuration of the mixing 
layer in PNe is also expected to occur within wind-blown bubbles that 
exhibit diffuse, soft X-ray emission such as the Orion Nebula, 
Wolf-Rayet bubbles, and superbubbles 
\citep[e.g.,][]{Gudel2008,Jaskot2011,ruiz13,Toala2012}.  Future UV 
observations towards other PNe and wind-blown bubbles, such as the 
Wolf-Rayet bubble NGC\,6888, will help us understand and unveil the 
physics of the mixing layer and its relation to the existence of 
the diffuse X-ray-emitting gas in hot bubbles.

\acknowledgments

Support for the \emph{Hubble Space Telescope} Cycle 20 General 
Observer Program 12509 was provided by NASA through grant 
HST-GO-12509.01-A from the Space Telescope Science Institute, 
which is operated by the Association of Universities for Research 
in Astronomy, Inc., under NASA contract NAS\,5-26555. 
X.F., M.A.G., and J.A.T.\ are partially funded by grant 
AYA~2011-29754-C03-02 of the Spanish MEC (Ministerio de Econom\'\i a 
y Competitividad) cofunded with FEDER funds. 
We thank Dr. Yong Zhang at the University of Hong Kong for discussion. 
We also thank the anonymous referee, whose comments help to enhance 
this paper. 
This paper utilizes the image from \emph{Chandra} X-ray Center 
(http://chandra.harvard.edu/), which was operated for NASA by the 
Smithsonian Astrophysical Observatory and developed with funding 
from NASA under contract NAS8-03060. 
This research has made use of NASA's Astrophysics Data System 
(http://adsabs.harvard.edu) and the Mikulski Archive for Space 
Telescope (MAST, https://archive.stsci.edu/). 
\\

{\it Facilities:} \facility{\emph{HST} (STIS)}.

\end{document}